# SHAPE-INDUCED MAGNETIC ANISOTROPY IN DILUTE MAGNETIC ALLOYS

V. N. GLADILIN[1], V. M. FOMIN[1], J. T. DEVREESE[2]
*Theoretische Fysica van de Vaste Stof, Universiteit Antwerpen (U.I.A.), Universiteitsplein 1, B-2610 Antwerpen (Belgium)*

Abstract:   We extend the theory of the surface-induced magnetic anisotropy to mesoscopic samples with arbitrary geometry. The shape-induced anisotropy of impurity spins in small brick-shaped grains of dilute magnetic alloys is studied in detail. The surface-induced blocking of a magnetic-impurity spin is shown to be very sensitive to geometric parameters of a grain. This implies that the apparent discrepancy between the experimental data of different groups on the size dependence of the Kondo resistivity can result from different microstructure of the used samples. In order to interpret recent experimental data on the anomalous Hall effect in thin polycrystalline Fe doped Au films, we analyse the magnetisation of impurity spins as a function of the impurity position and of the grain shape.

## 1.   INTRODUCTION

Recently, with the opportunities provided by nanolithography, finite-size effects in dilute magnetic alloys have been the subject of extensive experimental studies. While some experiments [1-3] revealed a considerable decrease of the logarithmic Kondo anomaly in the resistivity of thin films and narrow wires already for the µm scale, other measurements [4] showed an almost constant Kondo anomaly for wires with width ranging down to 40 nm. The apparent discrepancy between the results of different groups makes the subject controversial and far from conclusive [5].

---

[1] Permanent address: Fizica Structurilor Multistratificate, Universitatea de Stat din Moldova, str. A. Mateevici 60, MD-2009 Chisinau (Moldova).
[2] Also at: Universiteit Antwerpen (RUCA), Groenenborgerlaan 171, B-2020 Antwerpen, Belgium and Technische Universiteit Eindhoven, P.O. Box 513, 5600 NB Eindhoven, The Netherlands.

In the limit of strong disorder, the size dependence of the Kondo effect was investigated using the theory of weak localisation [6]. In the opposite case of ballistic regime, the size effects in dilute magnetic alloys can be linked to the surface-induced anisotropy of the magnetic-impurity spins [7,8]. The theory formulated in Refs. [7,8] for a semi-infinite sample has been generalised in Ref. [9] for mesoscopic wires with rectangular cross-section. Here we derive the Hamiltonian, which allows one to describe the surface-induced anisotropy in samples of various shapes. In order to investigate the influence of a polycrystalline structure of the samples on the anisotropy effect, we analyse the impurity-spin ordering in mesoscopic grains. Based on the developed model, we calculate the shape-dependent magnetisation of impurities in those grains. The obtained results are applied to interpret the recent experimental data on the anomalous Hall effect in thin polycrystalline AuFe films [10].

## 2. SURFACE-INDUCED MAGNETIC ANISOTROPY IN MESOSCOPIC SAMPLES

In mesoscopic samples of dilute magnetic alloys, the interaction of a magnetic impurity with the conduction electrons, which undergo spin-orbit scattering from the nonmagnetic host atoms, can result in the shape-induced anisotropy for the impurity spin [7-9]. The Hamiltonian of electrons, which interact with the magnetic impurity, can be written [11] as

$$H_0 = \sum_{k,m,s} \varepsilon_k a^+_{kms} a_{kms} + \sum_{k,m,m',s,s'} J \varepsilon_k a^+_{kms} (S\sigma)_{ss'} a_{km's'} d_{mm'}, \qquad (1)$$

where $J$ is the effective Kondo coupling constant. $a^+_{kms}$ ($a_{kms}$) creates (annihilates) a conduction electron with momentum $k$, angular momentum $l$ and its $z$-component $m$. $s$ is the spin quantum number, $\sigma$ is the Pauli-matrix vector of the electron. Since only $d$ channels of the electron-impurity interaction are considered, the index $l$ is taken to be equal to 2 and is dropped henceforth. The Hamiltonian of the host-atom $d$ orbitals is [7,8]

$$H_1 = \sum_{n,m,s} e_0 b_{ms}^{(n)+} b_{ms}^{(n)} + \sum_{n,m,m',s,s'} \lambda b_{ms}^{(n)+} (\mathbf{L}\sigma)_{ms,m's'} b_{ms'}^{(n)}$$
$$+ \sum_{n,m,m',s,s'} \left[ V_{kmm'}(\mathbf{R}_n) b_{ms}^{(n)+} a_{kms'} + \text{H.c.} \right], \quad (2)$$

where $b_{ms}^{(n)+}$ ($b_{ms}^{(n)}$) creates (annihilates) the host atom orbital at the site $n$, determined by the radius-vector $\mathbf{R}_n$ in the frame of reference with the origin at the magnetic impurity. $\lambda$ is the strength of the spin-orbit coupling. L is the orbital angular momentum of the host atom at the site $n$ and $V_{kmm'}(\mathbf{R}_n)$ is the Anderson localisation matrix element.

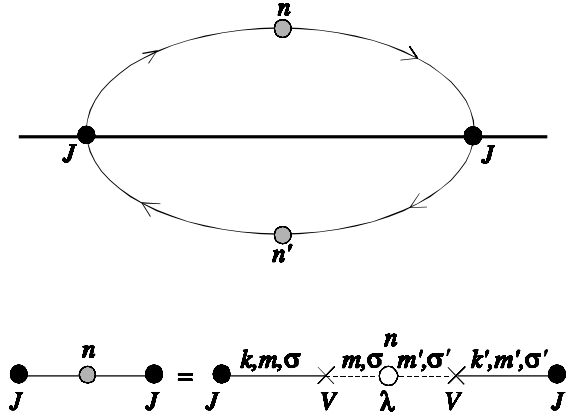

*Figure 1.* The self-energy Feynman diagram relevant to the impurity-spin anisotropy. The impurity spin, the conduction electron and the host-atom orbital are represented by the heavy, thin and dashed lines, respectively. The electron-impurity spin interaction, the Anderson hybridisation and the spin-orbit interaction are shown, respectively, by the filled circles, the crosses and the open circles. The grey circles stand for the host atoms at the sites $n$ and $n'$.

Both the electron-impurity interaction and the spin-orbit interaction are treated as perturbations. To the lowest order in these perturbations, the anisotropic contribution to the self-energy of the impurity spin arises from the Feynman diagram shown in Fig. 1. This diagram takes into account the interaction of the conduction electron with two host atoms at the sites $n$ and $n'$. The Hamiltonian $H_{\text{an}}$, which describes the impurity-spin anisotropy, can be obtained by integrating the contribution $F(\mathbf{R}_n, \mathbf{R}_{n'})$ of the diagram depicted in Fig. 1 over the host-atom positions:

$$H_{\text{an}} = \frac{1}{a^6} \int_{\text{sample}} d^3 R \int_{\text{sample}} d^3 R' F(\mathbf{R}, \mathbf{R}'), \quad (3)$$

where $a^3$ is the volume per host atom. The integral of Eq. (3) can be represented as

$$\int_{\text{sample}} d^3R \int_{\text{sample}} d^3R' F(\mathbf{R},\mathbf{R}') = \int_{\text{whole space}} d^3R \int_{\text{whole space}} d^3R' F(\mathbf{R},\mathbf{R}')$$
$$+ \int_{\text{outer domain}} d^3R \int_{\text{outer domain}} d^3R' F(\mathbf{R},\mathbf{R}') - 2 \int_{\text{outer domain}} d^3R \int_{\text{whole space}} d^3R' F(\mathbf{R},\mathbf{R}'). \qquad (4)$$

It is obvious that the first term in the right-hand side of Eq. (4) gives no anisotropic contribution. As shown in Ref. [8], the contribution $F(\mathbf{R},\mathbf{R}')$ is a rapidly decreasing function of the distances between the impurity and host atoms. Therefore, assuming that the minimal distance $d_{\min}$ between the impurity and the surface of the sample is much larger than $1/k_F$ ($k_F$ is the Fermi wavenumber), the second term in the right-hand side of Eq. (4) can be neglected with respect to the last term. Note that the integral over $\mathbf{R}'$ in the latter term does not depend on the shape of a sample. Using for $F(\mathbf{R},\mathbf{R}')$ the explicit expression from Ref. [8] and performing integration over $\mathbf{R}'$ and $R$, we obtain [to leading order in $1/(k_F d_{\min})$]

$$H_{\text{an}} = \frac{4\mathbb{A}}{\pi k_F} \int_0^{2\pi} d\varphi \int_0^{\pi} d\vartheta \sin\vartheta \frac{(\mathbf{S}\cdot\mathbf{e}(\vartheta,\varphi))^2}{R_s(\vartheta,\varphi)}, \qquad (5)$$

where $R = R_s(\vartheta,\varphi)$ is the equation of the surface of a sample and $\mathbf{e}(\vartheta,\varphi)$ is the unit vector in the direction determined by the angles $\vartheta$ and $\varphi$. The material-dependent constant $\mathbb{A}$ ranges between 0.01 and 1 eV for dilute AuFe alloys [8,9]. Integration over angles in Eq. (5) can be transformed into integration over the surface of a sample:

$$H_{\text{an}} = \frac{4\mathbb{A}}{\pi k_F} \int_{\text{surface}} d\mathbf{s}\cdot\mathbf{R}_s \frac{(\mathbf{R}_s\cdot\mathbf{S})^2}{R_s^6}, \qquad (6)$$

where $\mathbf{R}_s$ is the radius-vector of the surface element $d\mathbf{s}$ in the frame of reference with the origin at the magnetic impurity.

## 3.    MAGNETIC ANISOTROPY IN A BRICK-SHAPED GRAIN

Applying the expression (6) to a brick-shaped grain, the Hamiltonian $H_{\text{an}}$, which describes the surface-induced anisotropy, is obtained in the form

$$H_{an} = \mathbb{A}\left[S_x^2 b(x,y,z) + S_y^2 b(y,x,z) + S_z^2 b(z,x,y) + (S_x S_y + S_y S_x) c(x,y,z)\right.$$
$$\left. + (S_x S_z + S_z S_x) c(x,z,y) + (S_y S_z + S_z S_y) c(y,z,x)\right]. \tag{7}$$

$S_a$ ($a = x, y, z$) are the operators for the components of the impurity spin **S**. In the frame of reference with the origin at the center of the grain, the dependence of the anisotropy on the impurity position ($x,y,z$) is described by the functions

$$b(\mathbf{a},\mathbf{b},\mathbf{g}) = \sum_{l,m,n=\pm 1} u(\mathbf{a}_l, \mathbf{b}_m, \mathbf{g}_n), \tag{8}$$

$$c(\mathbf{a},\mathbf{b},\mathbf{g}) = -\sum_{l,m,n=\pm 1} lm\, v(\mathbf{a}_l, \mathbf{b}_m, \mathbf{g}_n), \tag{9}$$

where

$$\mathbf{a}_l = |\mathbf{a} + l\mathbf{a}_\mathbf{a}/2|, \tag{11}$$

$$u(p,q,t) = \frac{q}{p} v(p,q,t) + \frac{t}{p} v(p,t,q), \tag{12}$$

$$v(p,q,t) = \frac{1}{2p k_F \sqrt{p^2 + q^2}} \arctg\left(\frac{t}{\sqrt{p^2 + q^2}}\right), \tag{13}$$

$a_x$, $a_y$, $a_z$ are the dimensions of the grain.

In grains, the presence of differently oriented surfaces leads to a rather intricate behaviour of the impurity-spin anisotropy as compared to the case of a film considered in Refs. [7,8]. This is illustrated by Fig. 2, which shows the surface-induced splitting of the impurity-spin energy levels in AuFe grains. It is worth noting that, at any value of the constant $\mathbb{A}$, there exist specific regions within a grain, where the lowest state of an impurity spin is degenerate or quasi-degenerate. The impurities located in those regions can contribute to the Kondo effect even at relatively low temperatures. Such a partial (or even complete – depending on the position of an impurity and on the shape of a grain) cancellation of the anisotropy effect is a qualitatively new feature as compared to the case of a thin single-crystal AuFe film. In a film, the non-degenerate ground state is separated from other states by an energy interval $D > \mathbb{A}/t$ ($t$ is the film thickness) and, therefore, all the impurities with an integer spin become paramagnetic at temperatures $k_B T < D$ ($k_B$ is the Boltzmann constant) [7,8]. As seen from Fig. 2, the surface-induced magnetic anisotropy for impurity spins is very sensitive to the shape and size of grains. This implies that the apparent discrepancy between various experimental results [1-4] for the size dependence of the Kondo resistivity can be linked to different microstructures of the samples.

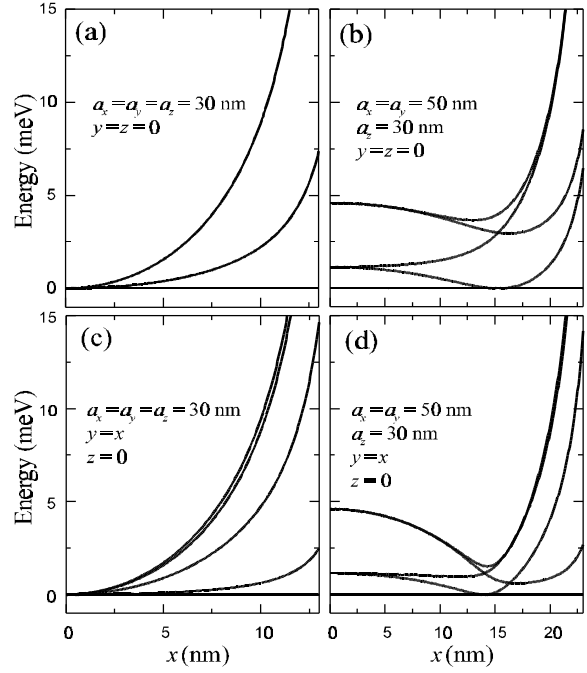

*Figure 2.* Energy spectra of an impurity spin (*S*=2) for various impurity positions in a cubic grain [panels (a) and (c)] and in a flat parallelepiped grain [panels (b) and (d)] at $\Lambda = 0.12$ eV. The energy of the lowest state of the impurity spin is taken to be zero.

## 4. IMPURITY-SPIN MAGNETISATION

The position-dependent splitting of the impurity-spin energy levels due to the shape-induced anisotropy can reveal itself through a substantially inhomogeneous response to an applied magnetic field. In a magnetic field ***B*** parallel to the *z*-axis, the magnetisation of an Fe spin is given by

$$\langle S_z \rangle = -\frac{1}{2\boldsymbol{m}_{\mathrm{B}} Z} \sum_{k=1}^{5} \exp\left(-\frac{E_k}{k_{\mathrm{B}}T}\right)\frac{\mathrm{d}E_k}{\mathrm{d}B}, \qquad (14)$$

where $\boldsymbol{m}_{\mathrm{B}}$ is the Bohr magneton, $Z = \sum_{k=1}^{5} \exp(-E_k/k_{\mathrm{B}}T)$ is the partition function for the impurity spin, and *T* is the temperature. The index $k = 1,\ldots,5$ labels the roots $E_k$ of the secular equation

$$\left| H_{S'_z S_z} - E d_{S'_z S_z} \right| = 0 \quad (S'_z, S_z = -2, -1, 0, 1, 2) \tag{15}$$

with the Hamiltonian $H = -2m_B S_z B + H_{an}$. The energy spectra of an impurity spin, which are shown in Fig. 3 as a function of magnetic field, appear to be qualitatively different for different positions of the impurity. Side by side with the energy spectra typical for magnetic impurities in a thin film subjected to a perpendicular magnetic field [see Fig. 3(a)], there are spectra similar to those in bulk [Fig. 3(b)] as well as spectra, which have no analogue in films or in bulk [Figs. 3(c) and 3(d)].

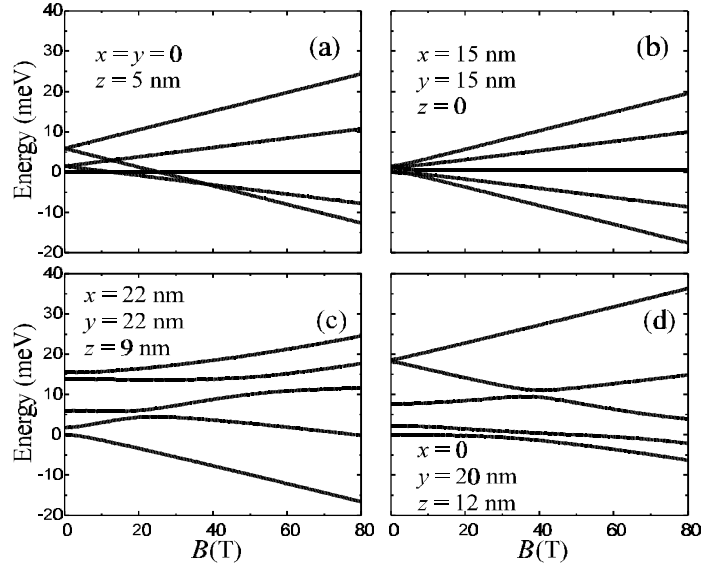

*Figure 3.* Energy spectra of an impurity spin ($S=2$) as a function of magnetic field parallel to the $z$-axis are shown for various impurity positions in a grain with $a_z = 30$ nm and lateral size $a = a_x = a_y = 50$ nm at $\Delta = 0.12$ eV. The energy of the lowest state of the impurity spin at $B = 0$ is taken to be zero.

As demonstrated by Fig. 4, the co-existence of impurity spins with substantially different energy spectra results in a strongly pronounced inhomogeneity of magnetisation within a grain. In a flat grain, the differential magnetisation at weak magnetic fields is significant only for impurities located near the grain edges parallel to the magnetic field, where the surface-induced magnetic anisotropy is appreciably weakened due to the competitive influence of perpendicular to each other surfaces. With increasing magnetic field, the magnetisation of those impurities rapidly saturates, while the magnetic response of the impurities in the central region of the grain reveals itself only at relatively high magnetic field.

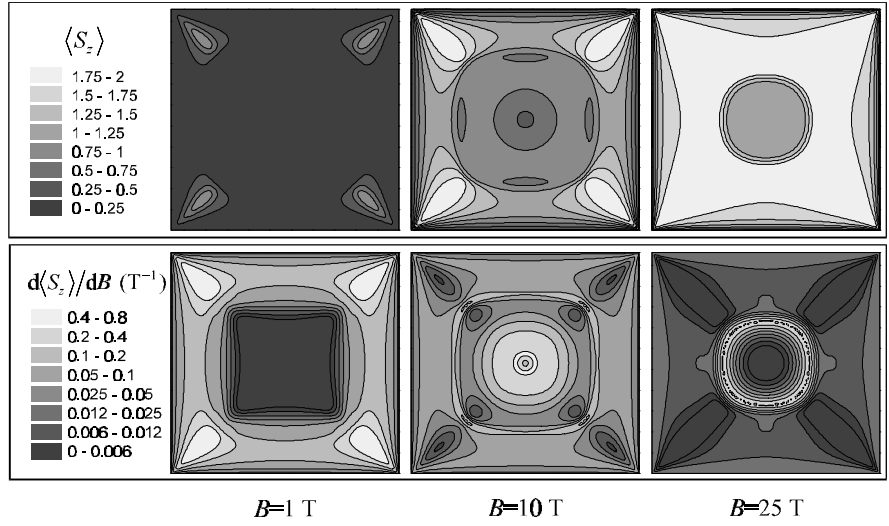

*Figure 4.* Spatial distributions of the magnetisation (upper row of panels) and of the differential magnetisation (lower row of panels) for the cross-section $z = 0$ of a grain with $a_z = 30$ nm and $a = a_x = a_y = 50$ nm at $\Delta = 0.12$ eV and $T = 1$ K.

In Fig. 5(a) the calculated differential magnetization averaged over impurity positions within a grain, $[d\langle S_z \rangle /dB]_{gr}$, is shown as a function of the magnetic field for different lateral dimensions $a$ of the grains. In the limiting case of a single-crystal film ($a \to \infty$), the impurity spin states at $B = 0$ are known [7,8] to be the eigenstates of $S_z$ with the energy eigenvalues proportional to $S_z^2$. At low temperatures, only the state with $S_z = 0$ is populated. Hence, the impurity spin does not respond to a weak magnetic field. When increasing $B$, the energy level with $S_z = 1$ becomes lower then that with $S_z = 0$. This gives rise to the first peak in the differential magnetization as a function of magnetic field. The second peak appears when the state with $S_z = 2$ becomes the ground state. Contrary to the case of a film, in grains, the presence of impurities with bulk-like behaviour can result in a relatively high response at weak magnetic fields.

When lateral sizes $a$ of a grain are larger than (but still comparable to) its height $a_z$, our model predicts that a minimum of $[d\langle S_z \rangle /dB]_{gr}$ can appear at moderate magnetic fields. For these grains the initial part of the calculated curves 'differential magnetisation versus magnetic field' is similar to the curves of the differential Hall resistivity versus $B$, measured for granular AuFe films [see Fig. 5(b)]. As shown in Fig. 6(a), due to an inhomogeneity of the shape-induced magnetic anisotropy, the non-linear behaviour of the impurity-spin magnetisation at low fields becomes less pronounced when moving the impurity towards the surface. This is consistent with the experimental data [10] for Au/AuFe/Au and AuFe/Au/AuFe trilayers [see Fig. 6(b)]. Of course, the experimental samples contain grains of various shape and size and a detailed fitting of the experimental curves requires an averaging over an ensemble of grains. The results of such a fitting are reported elsewhere [10].

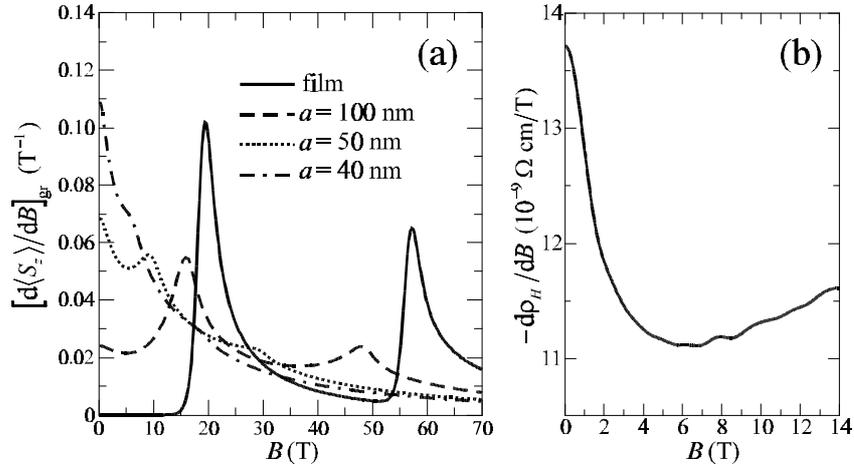

*Figure 5.* (a) Magnetic field dependence of the averaged differential magnetisation for grains with height $a_z = 30$ nm and various lateral dimensions $a$ at $\Delta = 0.12$ eV and $T = 1$ K. (b) Differential Hall resistivity as a function of magnetic field for a 2 at.% AuFe film with thickness 30 nm at $T = 1$ K [10].

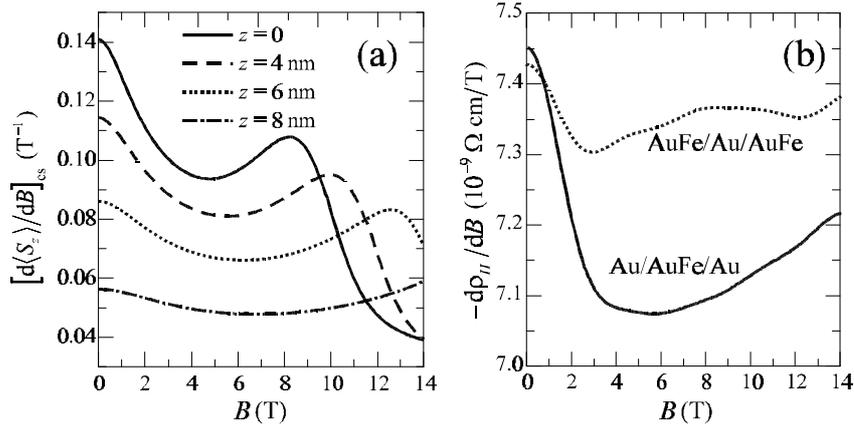

*Figure 6.* (a) Calculated magnetic field dependence of the differential magnetisation for a grain after averaging over a cross-section perpendicular to the magnetic field. Different curves are for different distances $z$ to the center of the grain with $a_z = 30$ nm and $a = 50$ nm ($\Delta = 0.12$ eV, $T = 1$ K). (b) Magnetic field dependence of the differential Hall resistivity for the Au/AuFe/Au and AuFe/Au/AuFe trilayers (Fe concentration is 3.5 at.% for the AuFe; the layer thickness is 15 nm for the central layer and 7.5 nm for the outer layers; $T = 1$ K) [10].

## 5. CONCLUSIONS

In small grains of dilute magnetic alloys, the competitive influence of differently oriented surfaces is shown to lead to a strong inhomogeneity of the surface-induced magnetic anisotropy for impurity spins. A high sensitivity of the anisotropy effect to the shape and size of grains implies that the apparent discrepancy between the experimental data of different groups [1-4] for the size dependence of the Kondo resistivity can be attributed to a difference in the microstructure of the samples. Our model provides an explanation for the experimentally observed suppression of the anomalous Hall resistivity in thin polycrystalline AuFe films at low magnetic fields as well as for the appearance of a minimum in the differential Hall resistivity at higher fields [10]. The results of our calculation are consistent with the observed dependence of the anomalous Hall resistivity on the location of magnetic impurities.

## ACKNOWLEDGEMENTS

This work has been performed in collaboration with E. Seynaeve, K. Temst, F. G. Aliev, and C. Van Haesendonck (Katholieke Universiteit Leuven, Belgium). It has been supported by the I.U.A.P., GOA BOF UA 2000, F.W.O.-V. projects Nos. G.0287.95, 9.0193.97 and the W.O.G. WO.025.99N (Belgium).